\documentclass[useAMS,usenatbib]{mn2e}

\usepackage{graphicx}
\long\def\symbolfootnote[#1]#2{\begingroup%
\def\thefootnote{\fnsymbol{footnote}}\footnote[#1]{#2}\endgroup}

\title[NUV--IR colours of red sequence galaxies]{NUV--IR colours of red sequence galaxies in local clusters}
\author[T. D. Rawle et al.]{Timothy D. Rawle$^{1}$\thanks{E-mail:
t.d.rawle@dur.ac.uk}, Russell J. Smith$^{1}$, John R. Lucey$^{1}$, Michael J. Hudson$^{2}$,\newauthor
Gary A. Wegner$^{3}$\\
$^{1}$Department of Physics, Durham University, Durham DH1 3LE, United Kingdom\\
$^{2}$Department of Physics and Astronomy, University of Waterloo, Waterloo, Ontario, N2L 3G1, Canada\\
$^{3}$Department of Physics and Astronomy, Dartmouth College, 6127 Wilder Laboratory, Hanover, NH 03755}
\begin{document}

\date{14 January 2008}

\pagerange{\pageref{firstpage}--\pageref{lastpage}} \pubyear{2007}

\maketitle

\label{firstpage}

\begin{abstract}
We present {\em GALEX} near-UV ($NUV$) and 2MASS $J$ band photometry for red sequence galaxies in local clusters. We define quiescent samples according to a strict emission threshold, removing galaxies with very recent star formation. We analyse the $NUV$--$J$ colour--magnitude relation (CMR) and find that the intrinsic scatter is an order of magnitude larger than for the analogous optical CMR ($\sim$0.35 rather than 0.05 mag), in agreement with previous studies. Comparing the $NUV$--$J$ colours with spectroscopically-derived stellar population parameters, we find a strong ($> 5.5 \sigma$) correlation with metallicity, only a marginal trend with age, and no correlation with the $\alpha$/Fe ratio. We explore the origin of the large scatter and conclude that neither aperture effects nor the UV upturn phenomenon contribute significantly. We show that the scatter could be attributed to simple `frosting' by either a young or a low metallicity subpopulation.
\end{abstract}


\section{Introduction}
\label{sec:intro}
The optical colour--magnitude relation (CMR) shows that brighter early-type galaxies are also redder \citep{vis77}, and is traditionally regarded as arising from the mass--metallicity sequence (cf. \citealt{dre84}, \citealt{kod97}). Gas loss, caused by supernova wind, occurs earlier in less massive galaxies. Therefore, a smaller fraction of gas is processed before being expelled from the less massive galaxy \citep{mat71}, resulting in lower average metallicities \citep{lar74}. \citet{bow92} found a very small intrinsic scatter in the $U$--$V$ CMR ($\sim$0.05 mag) and, due to the sensitivity of the $U$ band to the presence of young stars, interpreted this as a small age dispersion. Age and metallicity are observed to have the same effect on broadband optical colours, whereas spectral line indices can be used to break the degeneracy \citep{wor94}. \citet{kun98} claimed that the CMR is driven by metallicity variations with luminosity, although \citet{nel05} found evidence for a strong age--mass relation in addition to this metallicity--mass trend (see also \citealt{cal03}, \citealt{tho05}).

The ultraviolet--optical CMR for non-star-forming galaxies has an intrinsic scatter an order of magnitude larger than its optical counterpart; $\sim$0.5 mag compared to 0.05 mag (e.g. \citealt{yi05}). Hot young stellar populations dominate the ultraviolet (UV) flux for $\sim$100 Myr after an episode of star formation (ten times longer than H$\alpha$ emission after star formation; \citealt{lei99}). The large intrinsic scatter in the UV CMR is therefore often interpreted as differing quantities of very recent, albeit low level, star formation \citep{fer00}.

In intermediate age populations ($\sim$1--3 Gyr), the near-UV (NUV; 2000--3000\AA) flux is dominated by hot stars on the main sequence turn-off \citep[e.g.][]{oco99}. The sensitivity of the turn-off to the epoch of formation emphasises the importance of the UV bands for age determination \citep{dor03}.

Old ($\sim$10 Gyr) metal-poor populations have a significant UV flux contribution from very hot ($T_{\rm eff}$ $\sim$ 10000K) blue horizontal branch (BHB) stars \citep{mara00, lee02}. However these tend to reside in globular clusters or galactic haloes (where Fe/H $< -1$), where they are useful age indicators \citep{kav07}, rather than in relatively metal-rich elliptical galaxies.

The UV picture is further complicated by the presence of the ultraviolet upturn (or UV excess, UVX) phenomenon. First observed by \citet{cod69}, this unanticipated upturn dominates the far-UV (FUV; $<$ 2000\AA) in UVX galaxies. In contrast, the NUV can be decomposed into two separate components: the blue end of the main sequence/subgiant branch, and the UVX contribution \citep{dor97}. \citet{bur88} found that the UVX can sometimes be appreciable at wavelengths as long as 2700\AA: for example, in NGC 4649 $\sim$75 per cent of the NUV flux can be attributed to the UVX component. However, the UVX cannot be explained by the BHB population, as the temperature required to fit the upturn would be $T_{\rm eff}$ $\ga$ 20000K, whereas BHBs are usually no hotter than $T_{\rm eff}$ $\sim$ 12000K \citep{oco99}.

\citeauthor{bur88} further reported that FUV flux (assumed to trace the UVX) is strongly correlated with the Mg$_2$ line strength ($\sim$metallicity) and also with the velocity dispersion, which is a proxy for galaxy mass. However, more recent studies (e.g. \citealt{ric05}) have weakened the case for a strong UVX vs metallicity relation. From analysis of internal colour gradients, \citet{oco92} concluded that the FUV flux in most early-types originates from old stellar components. Drawing on these results, the source of the UVX is tentatively identified as hot, low mass, helium burning stars, such as extreme horizontal branch (EHB) or `failed' AGB (AGB-manqu\'{e}) stars and their progeny (see \citealt{yi97}, or the review \citealt{oco99}).

The {\em Galaxy Evolution Explorer} ({\em GALEX}; launched in 2003; \citealt{mar05}, \citealt{mor07}) is revolutionising UV astronomy, with high resolution imaging in two bands: near-ultraviolet ($NUV$; $\lambda_{eff} = 2310$ \AA) and far-ultraviolet ($FUV$; $\lambda_{eff} = 1530$ \AA). Using analysis of both $NUV$--$V$ and $FUV$--$V$ vs $B$--$V$ relations, \citet{don06} suggest that the $FUV$--$NUV$ colour reflects an extension of the colour--metallicity relation into the UV, as well as deducing that $\sim$10 per cent of ellipticals have residual star formation. Using the $NUV$--$\,r$ colour, \citet{kav06} also find non-negligible young stellar populations in morphologically selected early-type galaxies. \citet{sal07} derive star formation rates (SFRs) from broadband photometry dominated by the UV, and find good agreement with SFRs deduced from spectroscopic indices (predominantly using H$\alpha$). However, they also confirm that some galaxies with no H$\alpha$ emission show signs of star formation in the UV bands and attribute this to post-starburst galaxies. 
 
Here, we build upon these previous studies by exploring the relationship between the $NUV$--$J$ colour and spectroscopic stellar population indicators for a sample of quiescent, red sequence galaxies in nearby clusters. This paper is organised as follows. Section \ref{sec:data} describes our two red sequence samples and associated 2MASS and {\em GALEX} datasets. The criteria used to remove galaxies with emission are described. In Section \ref{sec:results} we show that a large intrinsic scatter is found in the $NUV$--$J$ colours of these quiescent cluster galaxies. Metallicity is shown to be strongly correlated with the $NUV$--$J$ colour, although there remains a large residual scatter. Section \ref{sec:discuss} discusses possible explanations for this scatter, showing that morphological abnormalities, aperture bias and the UV upturn do not contribute significantly. We investigate simple `frosting' models with a low mass fraction of younger stars (or alternatively a low mass fraction population of low metallicity, blue horizontal branch stars), and show that these could account for the scatter. The uncertainties in the $NUV$ K-correction are also discussed. Our conclusions are given in Section \ref{sec:conclusion}.

\section{Data}
\label{sec:data}

We use two complementary samples of red sequence galaxies in local clusters: the first is explicitly red sequence selected by optical colour, and is a large sample, containing $\sim$10 times the number of galaxies; the second has the advantage of higher quality spectroscopy, and uses an emission line cut (Sec. \ref{sec:cuts}) to ensure a red sequence sample. \cite{smi07} Fig. 1 demonstrates that H$\alpha$ selection efficiently removes all galaxies bluer than the red sequence and is more restrictive than a cut on colour.

\subsection{Galaxy samples}

\subsubsection{NFPS sample}
The NOAO Fundamental Plane Survey (NFPS; \citealt{smi04}, \citealt{nel05}) is a study of X-ray selected clusters distributed over the whole sky and at redshifts between 0.015 $<$ z $<$ 0.072. More than 4500 galaxies lying within 1 Mpc of the centre of each cluster, and within 0.2 mag of the cluster red sequence on the $B$--$R$ CMR (see \citealt{smi04} Fig. 3), were observed spectroscopically. Of these, 3514 have redshift, velocity dispersion and spectral line strength measurements (from 2 arcsec diameter fibres).

\subsubsection{Shapley Supercluster (SSC) sample}
The second sample of galaxies concentrates on the core of the Shapley supercluster (SSC; Abell clusters A3556, A3558, A3562 at z $\approx$ 0.049). This sample consists of 541 galaxies selected from  NFPS imaging but to a deeper limit ($R$ $<$ 18; \citealt{smi07}). Follow-up spectroscopy for these targets were obtained using 2 arcsec diameter fibres, equating to 2 kpc at the distance of Shapley. A set of three non-redundant line indices were fit to the models of \citet{tho03,tho04} in order to estimate age, metallicity (Z/H) and $\alpha$-abundance ($\alpha$/Fe) for each galaxy. The primary tracer of age in this scheme is H$\gamma$F; for metallicity Fe5015 is used; Mgb5177 is the $\alpha$-abundance indicator. This method is described in detail in \citet{smi08}.

\subsection{{\em GALEX} and 2MASS data}

\begin{table*}
\begin{minipage}{178mm}
\centering
\caption{GALEX $NUV$ images. $cz$ (km\,s$^{-1}$) is cluster redshift in the CMB frame. A single exposure is used unless column six indicates the number of coadded images. Total exposure time for coadded images is given for $NUV$ and (where the image is available) $FUV$.} 
\label{tab:nuvtiles} 
\begin{tabular}{@{}lcrlrcrr} 
\hline 
{\em GALEX} image & Centre RA & \multicolumn{1}{c}{Centre Dec} & Cluster(s) &  \multicolumn{1}{c}{$cz_{\rm CMB}$} & Images & \multicolumn{2}{c}{$t_{\rm exp}$ (secs)} \\ 
& (J2000) & \multicolumn{1}{c}{(J2000)} & & (km\,s$^{-1}$) & coadded & $NUV$ & $FUV$ \\
\hline 
GI1\_004001\_A2734 & 00 11 21.6 & --28 51 00 & A2734 & 18249 & 3 & 3535 & \\
GI1\_067001\_UGC0568\_0003 & 00 55 08.9 & --01 02 47 & A0119 & 12958 & & 1556 & 3024 \\
MISDR1\_16976\_0422 & 01 14 29.3 & +15 01 46 & A0160A & 12794 & & 1444 & 1444 \\
GI1\_004002\_A0262\_0001 & 01 52 45.6 & +36 08 58 & A0262 & 4464 & & 1698 & 1698 \\
NGA\_NGC1058 & 02 43 26.6 & +36 25 39 & A0376 & 14371 & & 1265 & 1265 \\
GI1\_004003\_A3104\_0001 & 03 14 21.6 & --45 25 12 & A3104 & 21560 & & 1588 & 1588 \\
GI1\_004027\_A3158\_0001 & 03 42 57.6 & --53 37 48 & A3158 & 17542 & & 1026 & 887 \\
GI1\_004004\_A3266 & 04 31 24.0 & --61 26 24 & A3266 & 17713 & 2 & 1441 & 1441 \\
GI1\_004005\_A0548 & 05 46 40.0 & --25 37 21 & A0548A/B & 12439 & 4 & 3155 & \\
GI1\_004006\_A3376 & 06 01 43.2 & --39 59 24 & A3376 & 14016 & 2 & 2176 & 2176 \\
GI1\_004007\_A3389\_0002 & 06 21 57.7 & --64 57 35 & A3389 & 8075 & & 976 & \\
MISDR1\_24335\_0270 & 10 13 07.7 & --00 23 25 & A0957 & 13849 & & 1703 & 1703 \\
GI1\_004025\_A3528\_0001 & 12 53 57.6 & --29 13 48 & A3528A/B & 16764 & & 1697 & \\
GI1\_004008\_A1644\_0002 & 12 57 12.0 & --17 24 36 & A1644 & 14478 & & 1163 & \\
GI1\_009003\_HPJ1321m31\_0001 & 13 21 05.8 & --31 32 20 & A3556 & 14660 & & 1615 & \\
GI1\_004010\_A3556 & 13 25 26.1 & --31 36 07 & A3556, A3558 & 14660 & 2 & 1805 & \\
GI1\_004011\_A3558\_0001 & 13 27 57.6 & --31 30 00 & A3558 & 14660 & & 1676 & \\
MISDR1\_33707\_0586 & 14 42 46.3 & +03 39 11 & MKW8 & 8449 & & 1698 & 1698 \\
GI1\_004016\_A1991\_0001 & 14 54 31.1 & +18 38 23 & A1991 & 17741 & & 967 & \\
GI1\_004026\_A2063\_0001 & 15 23 36.0 & +08 36 34 & A2063 & 10444 & & 1512 & \\
NGA\_NGC6166 & 16 28 39.9 & +39 33 24 & A2199 & 8872 & & 1437 & 1437 \\
GI1\_004020\_A3716 & 20 51 57.5 & --52 46 48 & A3716 & 13141 & 2 & 3179 & 1689 \\
GI1\_004021\_A2399\_0004 & 21 57 19.1 & --07 47 59 & A2399 & 17046 & & 1322 & \\
GI1\_004022\_A2589 & 23 23 57.6 & +16 46 47 & A2589 & 12001 & 3 & 4252 & \\
GI1\_004023\_A4059\_0003 & 23 56 59.9 & --34 45 35 & A4059 & 14660 & & 635 & \\
\hline 
\end{tabular} 
\end{minipage}
\end{table*} 

{\em Galaxy Evolution Explorer} ({\em GALEX}) near-ultraviolet ($NUV$) band images are available for 26 (from a total of 93) NFPS clusters. Due to a detector fault, only some clusters have associated far-ultraviolet ($FUV$) band images. Most of the images are from our guest investigator snapshot programme ({\em GALEX} GI1\_004) which targeted a subset of NFPS clusters with low galactic extinction and having a large number of spectra from \citet{smi04}. In addition, a small number of images of comparable depth from the GR2 GI archive, medium imaging survey (MIS) and near galaxy survey (NGS) have been used. Table \ref{tab:nuvtiles} lists all the images analysed along with their centre position and stacked exposure time. {\em GALEX} images have a 1.25\degr\ diameter, but only the central 1.2\degr\ field has been analysed due to the poor image quality at the edges. {\em GALEX} images have a plate scale of 1.5 arcsec pixel$^{-1}$ and a PSF FWHM of $\sim$5 arcsec.

We analyse infrared tiles from the $J$ band ($\lambda_{eff} = 1.25 \mu$m) of the Two Micron All Sky Survey (2MASS; \citealt{skr06}). We measure directly from the tiles, rather than adopt photometry from the 2MASS extended source catalogue \citep[XSC;][]{jar00}, as some of our target objects are unresolved. Additionally, we require the PSF to match that of the NUV images (2MASS $J$ PSF FWHM $\sim$3 arcsec) and therefore Gaussian smooth the 2MASS tiles before analysis. 2MASS tiles have a plate scale of 1 arcsec pixel$^{-1}$.

{\sc SExtractor} \citep{ber96} was employed in dual image mode to detect all objects in the $J$ band. Photometry was measured for all sample targets in both bands using a range of matched apertures: a Kron-type aperture ({\sc SExtractor}'s MAG\_AUTO) and seven apertures 3--34.5 arcsec in diameter (MAG\_APER). Throughout this work, $J$ band Kron apertures are used for total luminosity, and matched 12 arcsec diameter apertures for $NUV$--$J$ colours. Our $J$ band photometry is in good agreement with 2MASS ($\sim$0.2 mag RMS) for objects in the XSC.

Targets with {\sc SExtractor} apertures flagged as truncated, or with a deblending error, have been removed from the sample, and only confirmed cluster members with redshift and log$\,\sigma$ measurements within the \citet{smi04} or \citet{smi07} datasets (NFPS and SSC respectively) are used in the analysis. Table \ref{tab:num_gals} lists the number of galaxies in the samples at this stage, and after subsequent restrictions.

\begin{table}
\centering
\caption{Size of the galaxy samples, following the selection criteria applied.} 
\label{tab:num_gals} 
\begin{tabular}{@{}lrr} 
\hline
& NFPS & SSC \\ 
\hline
Original sample & 4527 & 541 \\
...with usable $NUV$ and $J$ photometry & 1493 & 307 \\
...with cz and log$\,\sigma$ data & 990 & 267 \\
...after emission line cut & 920 & 156 \\
......after optical apparent magnitude cut$^1$ & 544 & 101 \\
......with stellar population parameters$^2$ & -- & 87 \\
......with $FUV$ and H$\gamma$F data$^3$ & 222 & -- \\
\hline 
\end{tabular}

NOTE -- see $^1$(Sec \ref{sec:compare}), $^2$(Sec \ref{sec:derived}) or $^3$(Sec \ref{sec:uvx})
\end{table} 

All colours and magnitudes are measured in the AB system ($J_{\rm AB}=J_{\rm Vega}+0.91$; \citealt{bla05}), and have been corrected for galactic extinction using the reddening maps of \citet{sch98}; A$_{FUV}$=8.29$\times$$E$($B$--$V$), A$_{NUV}$=8.87$\times$$E$($B$--$V$), A$_J$=0.902$\times$$E$($B$--$V$) (\citealt{car89}; Schlegel et al.).

Estimates of the K-correction in the UV bands are currently derived empirically from poorly constrained spectra, and therefore subject to large uncertainties. \citet[fig. 22]{kav06} estimate that the $NUV$--$\,i$ correction would be $\sim$0.1 mag throughout the range in redshift considered here. We do not apply a K-correction in this study. This issue is addressed further in Section \ref{sec:kcor}.

\subsection{Emission-line cuts}
\label{sec:cuts}

In order to construct a sample of emission-free red sequence galaxies, a restriction is made on emission line strengths. Mindful of the effect of nebular emission `fill in' for the age-sensitive Balmer lines (H$\beta$, H$\gamma$ and H$\delta$), the preferred cut is on the H$\alpha$ line. Unfortunately, H$\alpha$ was not measured for most NFPS galaxies, so the selection criteria of the original NFPS reduction has been adopted \citep{nel05}. Specifically, this involves a cut on the H$\beta$ equivalent width, {\em EW}(H$\beta$) $> -0.6$ \AA{ }(negative {\em EW} denotes spectral line emission) supplemented by a cut on OIII $\lambda$5007, {\em EW}(OIII $\lambda$5007) $> -0.8$ \AA.{ }The cut on the SSC sample, which has H$\alpha$ measurements, follows the prescription of \citet{smi07}, which uses {\em EW}(H$\alpha$) $> -0.5$ \AA{ }(approximately equivalent to {\em EW}(H$\beta$) $> -0.2$ \AA). The OIII cut would not remove any additional galaxies. These cuts ensure red sequence subsamples, free of galaxies with a sizeable star formation component or an optically strong active galactic nucleus. The data and photometry for the resulting subsamples are reported in Tables \ref{tab:data1} and \ref{tab:data2}.

\begin{table*}
\begin{minipage}{178mm}
\centering
\caption{Data for galaxies in the NFPS sample. Galaxy position is encoded in the ID. $cz_{\rm CMB}$ (km\,s$^{-1}$) is the mean cluster redshift in the CMB frame. $cz_{\rm hel}$ (km\,s$^{-1}$) is galaxy redshift in the heliocentric frame. Magnitudes (Kron-type apertures) and colours (12 arcsec diameter apertures) are in the AB system and have been galactic extinction corrected, but not K-corrected.} 
\label{tab:data1} 
\begin{tabular}{@{}lccccc} 
\hline
Galaxy ID & cluster & galaxy & apparent $J$ & $NUV$--$J$ & $FUV$--$J$ \\ 
& $cz_{\rm CMB}$ & $cz_{\rm hel}$ & & & \\
\hline
NFPJ043305.0--612235 & 17714 & 16144 & 15.750 $\pm$ 0.071 & 6.732 $\pm$ 0.142 & 7.205 $\pm$ 0.250 \\
NFPJ043306.7--612614 & 17714 & 17702 & 14.898 $\pm$ 0.047 & 7.091 $\pm$ 0.102 & 7.497 $\pm$ 0.184 \\
NFPJ043307.6--611338 & 17714 & 16356 & 14.659 $\pm$ 0.042 & 6.979 $\pm$ 0.095 & 7.659 $\pm$ 0.195 \\
NFPJ054415.7--255429 & 12939 & 10674 & 14.693 $\pm$ 0.045 & 6.886 $\pm$ 0.077 & - \\
NFPJ054431.6--255550 & 12939 & 13247 & 14.783 $\pm$ 0.047 & 6.767 $\pm$ 0.086 & - \\
\hline 
\end{tabular} 

Full content of this table is available in the electronic version
\end{minipage}
\end{table*} 

\begin{table*}
\begin{minipage}{178mm}
\centering
\caption{Data for galaxies in the SSC sample. As in Table \ref{tab:data1}. Supercluster mean $cz_{\rm CMB}$ = 14660 km\,s$^{-1}$ for all galaxies. Stellar population parameters given where available (age in Gyr).} 
\label{tab:data2} 
\begin{tabular}{@{}lcccccc} 
\hline
Galaxy ID & $cz_{\rm hel}$ & apparent $J$ & $NUV$--$J$ & log(age/Gyr) & [Z/H] & [$\alpha$/Fe] \\
\hline
NFPJ132418.2--314229 & 13948 & 14.197 $\pm$ 0.035 & 7.088 $\pm$ 0.085 & 0.94 $\pm$ 0.03 & 0.27 $\pm$ 0.02  & 0.24 $\pm$ 0.02 \\
NFPJ132423.0--313631 & 14642 & 14.847 $\pm$ 0.048 & 6.706 $\pm$ 0.103 & 0.89 $\pm$ 0.05 & 0.11 $\pm$ 0.03  & 0.15 $\pm$ 0.02 \\
NFPJ132425.9--314117 & 13888 & 14.380 $\pm$ 0.038 & 6.604 $\pm$ 0.079 & -- & --  & -- \\
NFPJ132426.5--315153 & 14922 & 14.676 $\pm$ 0.043 & 6.575 $\pm$ 0.081 & 0.95 $\pm$ 0.04 & 0.23 $\pm$ 0.03  & 0.24 $\pm$ 0.02 \\
2MASXJ13250387--3132449 & 14266 & 14.792 $\pm$ 0.049 & 6.626 $\pm$ 0.109 & -- & --  & -- \\
\hline 
\end{tabular}

Full content of this table is available in the electronic version
\end{minipage}
\end{table*} 

\begin{figure}
\includegraphics[viewport=20mm 50mm 163mm 177mm,height=84mm,angle=270,clip]{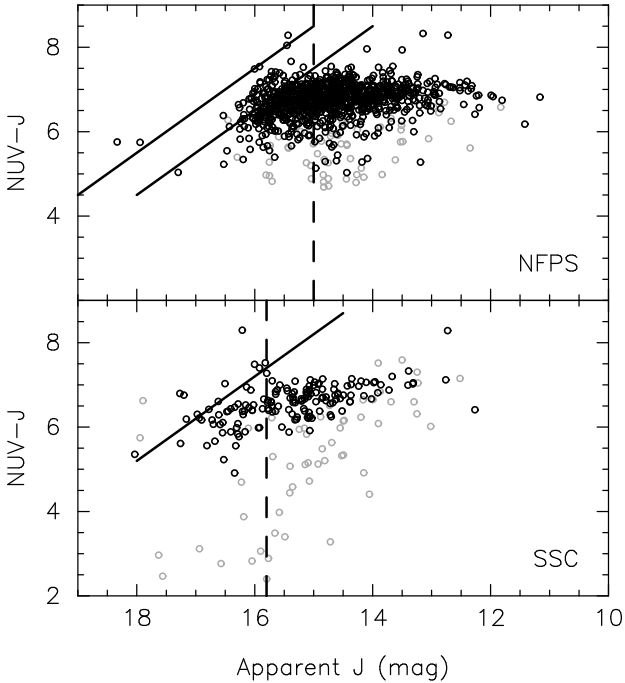}
\caption{Colour--magnitude diagram for the samples (NFPS upper panel; SSC lower panel). Grey points are those removed by emission line criteria detailed in Sec. \ref{sec:cuts}. The larger tail of grey points for SSC results from the lack of an explicit colour cut on the original sample. Solid lines are {\em GALEX} detection limits and dashed lines are the applied $J$ band apparent magnitude cuts. Median error bars are $\sim$0.05 mag in $J$ band and $\sim$0.1 mag in the colour.}
\label{fig:prelimcuts}
\end{figure}

\section{Results}
\label{sec:results}

\subsection{$NUV$--$J$ colour relations}
\label{sec:compare}

Figure \ref{fig:prelimcuts} shows the $NUV$--$J$ colour--(apparent) magnitude diagram for the NFPS and SSC samples. Shown in grey are the galaxies removed by the emission cut from Section \ref{sec:cuts}. Prior to the emission cut, SSC has a larger colour range than NFPS. This is because NFPS was explicitly selected on $B$--$R$ colour while SSC only on total $R$ band apparent magnitude. The H$\alpha$ cut efficiently removes the very blue objects.

All target galaxies were within the 2MASS $J$ band detection limit, but the $NUV$ band has a 5$\sigma$ detection limit of 22.5--23.5 mag, depending on the coadded image exposure time. Figure \ref{fig:prelimcuts} (upper panel) shows that for the NFPS, these limits result in a bias against faint red galaxies. Assuming the brightest $NUV$ detection limit, a sample cut is applied to the $J$ band apparent magnitude at 15.0 mag. For the SSC sample, all of the targets appear on just two $NUV$ images with similar exposure times, and a (5$\sigma$) detection limit of 23.2 mag. Although in practice only two SSC targets have a non-positive flux in the $NUV$ band (compared to $\sim$8 per cent in NFPS), for consistency, SSC has been treated in a similar manner, with a cut applied at 15.8 mag. These cuts are shown as dashed lines in Figure \ref{fig:prelimcuts} and ensure a complete sampling of the colour range over the selected luminosity interval.

\begin{figure}
\includegraphics[viewport=22mm 49mm 162mm 180mm,height=84mm,angle=270,clip]{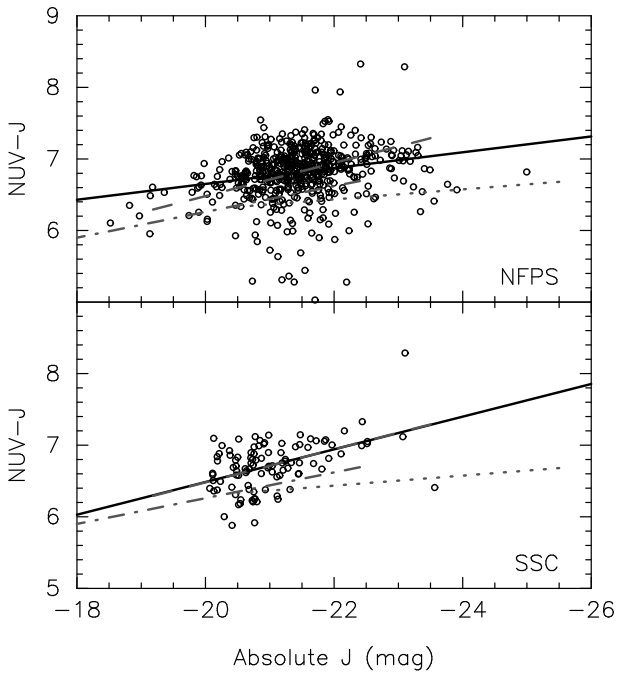}
\caption{$NUV$--$J$ colour--absolute magnitude diagram for both samples, with best fits in black. Relations from previous studies in grey: \citet{yi05}=dashed, \citet{bos05}=dotted, \citet{hai07}=dash-dotted. Median error bars are $\sim$0.05 mag in $J$ band and $\sim$0.1 mag in $NUV$--$J$.}
\label{fig:cmdiag}
\end{figure}

The colour--(absolute) magnitude diagrams for the low-emission galaxies are presented in Figure \ref{fig:cmdiag}. There is a correlation between the luminosity and colour; brighter galaxies tend to be redder. However, there is a large scatter; rms dispersions of 0.37 and 0.30 mag for NFPS and SSC samples respectively. The smaller scatter within the SSC sample is probably due to the slightly more restrictive Balmer emission line criteria (see Sec. \ref{sec:cuts}). The scatter in each sample does not increase by more than $\sim$10 per cent unless the cut criteria are relaxed beyond an equivalent width of -1 \AA. Only 5 per cent of the scatter can be accounted for by photometric measurement error. As intrinsic scatter dominates, all correlations in this study are computed without error weighting.

Table \ref{tab:bestfits} summarises the CMRs for our two samples. Additionally, the CMRs measured in three previous studies (\citealt{yi05}, \citealt{bos05}, \citealt{hai07}) are shown for comparison. Different sample selections were used for each of these studies. \citeauthor{yi05} use a `UV-weak' early-type galaxy sample, selected from SDSS by concentration index and luminosity profiles, and then by the flux ratios F($NUV$)/F($r$) and F($FUV$)/F($r$) both being less than 0.07. The sample covers a $J$ band luminosity range comparable to our work. \citeauthor{bos05} use a volume-limited sample of galaxies in the Virgo cluster, with a subsample defined as elliptical by visual classification. Haines et al. use a volume limited sample of local galaxies from SDSS, with the subsample (labelled `passive red sequence galaxies') restricted by the emission line criteria {\em EW}(H$\alpha$) $> -2$ \AA.

Despite these selection differences, our derived relations are in good agreement with the previous studies. The scatter is consistent ($\sim$0.3--0.5 mag) given the different sample definitions, and considerably large in comparison to that of the optical CMR ($\sim$0.05 mag; Bower et al. 1992).

\begin{table}
\begin{center}
\caption{$NUV$--$J$ CMRs.} 
\label{tab:bestfits} 
\begin{tabular}{@{}lccccc} 
\hline
& \multicolumn{2}{c}{original} & \multicolumn{2}{c}{transformed} & rms \\ 
& x & y & $a$ & $b$ & dispersion \\
\hline
NFPS & J & NUV--J & --0.11 & 4.45 & 0.37 \\
SSC & J & NUV--J & --0.23 & 1.93 & 0.30 \\
\hline
Y05$^1$ & r & NUV--r & --0.23 & 1.88 & 0.58 \\
B05$^2$ & H & NUV--H & --0.07 & 4.89 & 0.47 \\
H07$^3$ & r & NUV--r & --0.18 & 2.66 & 0.37 \\
\hline 
\end{tabular}
\end{center}
{\sc Notes} -- Original x and y parameters are given for reference. $a$ and $b$ are for relations in the form $NUV$--$J$ $= aJ + b$, assuming the following colours: ($J$--$H$)$_{\rm AB}$=0.2, $R_{\rm AB}$=$r$--0.21, ($J$--$R$)$_{\rm AB}$=$-$0.8. $^1$from \citet{yi05}. $^2$from \citet{bos05}. $^3$from \citet{hai07}.
\end{table} 

\begin{figure}
\includegraphics[viewport=22mm 49mm 162mm 180mm,height=84mm,angle=270,clip]{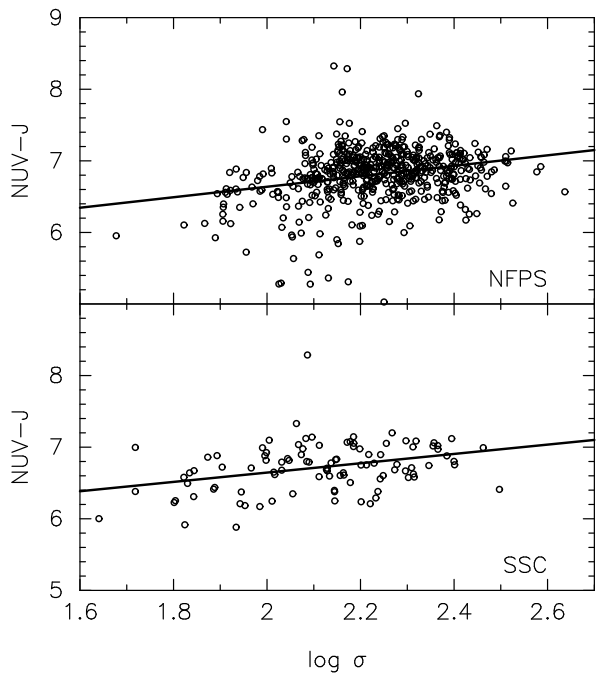}
\caption{$NUV$--$J$ vs log$\,\sigma$ for NFPS (upper panel) and SSC (lower panel). Median error bars are $\sim$0.02 and $\sim$0.01 for log$\,\sigma$ in the two samples respectively and $\sim$0.1 in the colour.}
\label{fig:logsig}
\end{figure}

Velocity dispersion provides an alternative mass proxy to luminosity, and, for optical colours, the $\sigma$ correlation appears more fundamental \citep{ber05}. Figure \ref{fig:logsig} presents the samples in terms of their $NUV$--$J$ colour and log$\,\sigma$. There is a clear correlation, with slopes of $0.73 \pm 0.11$ and $0.65 \pm 0.17$ for NFPS and SSC respectively, but the rms scatter (0.36 and 0.32 mags) is indistinguishable from that of the CMR.

\begin{figure}
\includegraphics[viewport=0mm 0mm 80mm 130mm,height=84mm,angle=270,clip]{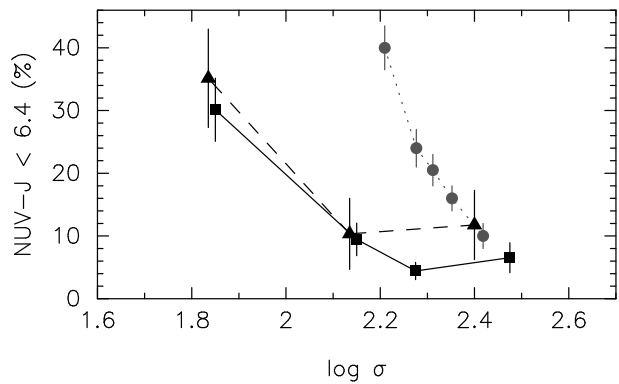}
\caption{Fraction of galaxies with $NUV$--$J$ $<$ 6.4 as a function of log$\,\sigma$. NFPS: squares/solid line; SSC: triangles/dashed line. Estimates from Figure 1 of \citet{sch06} are given for comparison (circles/dotted line) and show a much steeper increase in the number of blue galaxies with decreasing $\sigma$.}
\label{fig:blue_frac}
\end{figure}

Figure \ref{fig:blue_frac} shows the fraction of `blue' galaxies as a function of log$\,\sigma$ for our two samples. Blue galaxies are defined by $NUV$--$J$ $<$ 6.4 mag, as used in \citet{sch06} who study SDSS galaxies probing to much lower density environments than our samples. Their blue galaxy fractions are plotted in Fig. \ref{fig:blue_frac} and show a markedly higher fraction for a given velocity dispersion. For example, our two samples have a blue fraction of $\sim$40 per cent only at the lowest sigma (log$\,\sigma$ $<$ 1.8), while the \citeauthor{sch06} sample reaches this blue fraction at log$\,\sigma=2.2$. At face value, this result suggests a large difference between field and cluster galaxy populations. However, the differences in sample selection have to be considered (\citeauthor{sch06} use a sample selected on morphology), which is beyond the scope of this study.

\subsection{Stellar population parameters in the Shapley sample}
\label{sec:derived}

In order to investigate the physical origin of the large intrinsic scatter found in the $NUV$--$J$ colour, we examine the relationship between colour and the stellar population parameters (age, metallicity, $\alpha$-abundance) for the SSC sample.

In the following analysis, involving only the SSC sample, the derived limit on $J$ band apparent magnitude (Sec. \ref{sec:compare}) is not applied, allowing a larger number of galaxies. (It should be noted that the subsequent conclusions are robust against the use of the $J$ band cut.) The emission line selection criteria {\em are} retained, and the sample is further restricted by the overlap of the photometric and spectroscopic datasets (see Table \ref{tab:num_gals}).

\begin{figure}
\includegraphics[viewport=0mm 0mm 178mm 129mm,height=84mm,angle=270,clip]{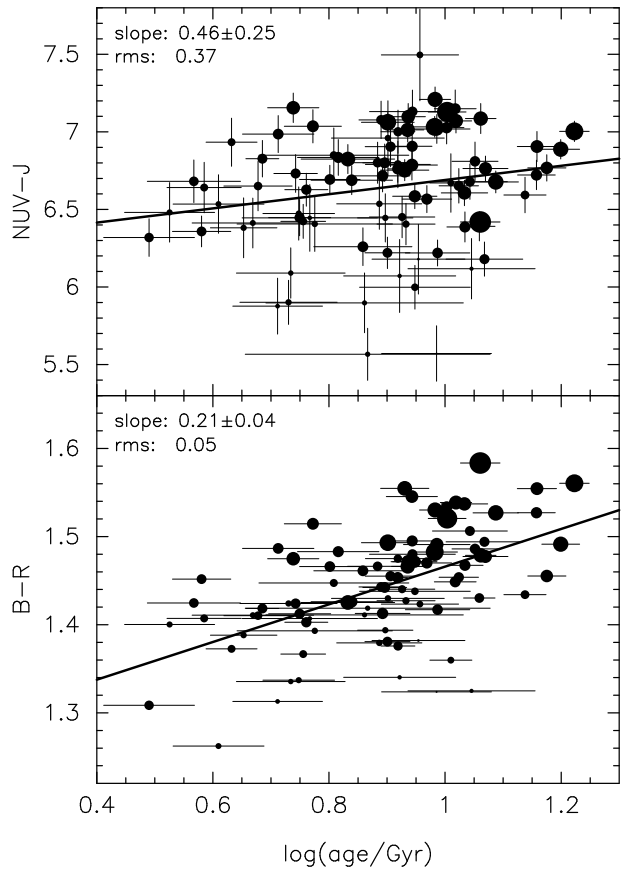}
\caption{$NUV$--$J$ (upper) and $B$--$R$ (lower; from \citeauthor{mer06}) colours vs log(age) for the SSC sample. Symbol size represents $J$ band luminosity; larger=brighter. Note that the lower panel has a much smaller range in the vertical scale.}
\label{fig:age}
\end{figure}

\begin{figure}
\includegraphics[viewport=49mm 0mm 148mm 129mm,height=84mm,angle=270,clip]{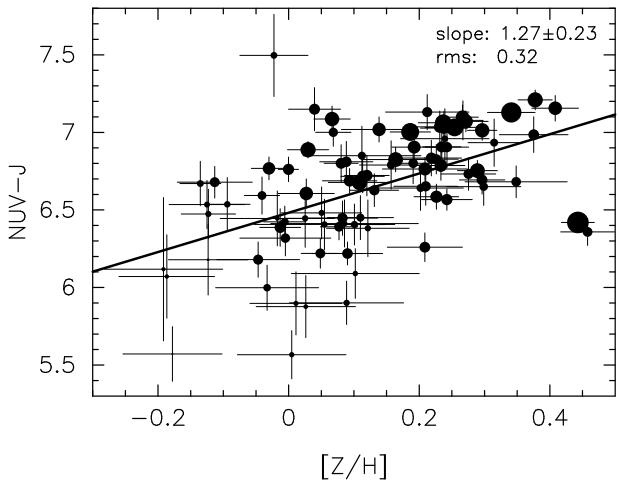}
\caption{$NUV$--$J$ vs metallicity [Z/H]. Symbol size represents $J$ band luminosity; larger=brighter.}
\label{fig:met}
\end{figure}

\begin{figure}
\includegraphics[viewport=50mm 0mm 148mm 129mm,height=84mm,angle=270,clip]{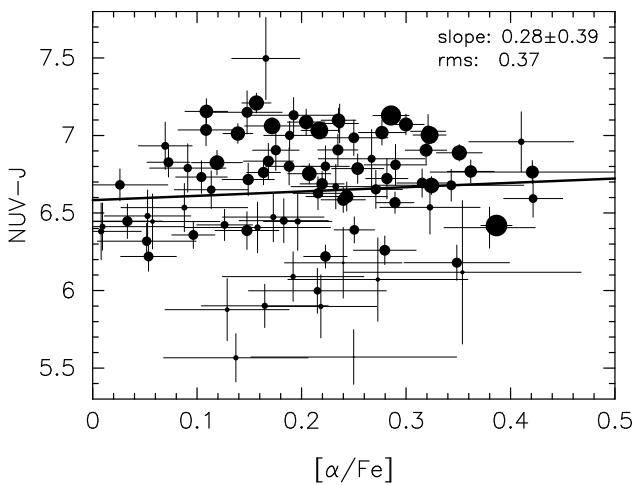}
\caption{$NUV$--$J$ vs $\alpha$-abundance [$\alpha$/Fe]. Symbol size represents $J$ band luminosity; larger=brighter.}
\label{fig:alphafe}
\end{figure}

$NUV$--$J$ vs log(age) is shown in Figure \ref{fig:age} (upper panel). Galaxies with nebular emission have been removed via the emission line criteria. Therefore, if the UV sources also contribute to the optical flux and have strong Balmer lines, the $NUV$--$J$ and age for the remaining objects would be correlated. We find only a marginal correlation between age and $NUV$--$J$, with a slope of $0.46 \pm 0.25$ and an intrinsic scatter (after accounting for the measurement error) of 0.33 mag. From the upper panel of Fig. \ref{fig:age}, it is also apparent that more luminous, and by inference larger, red sequence galaxies ($J$ band luminosity is shown in Figs. \ref{fig:age}--\ref{fig:alphafe} by the symbol size) are not solely confined to the redder $NUV$--$J$ colours, although they do tend to be the oldest.

Figure \ref{fig:age} (lower panel) shows the relation between age and $B$--$R$ colour for the same sample of galaxies (note the much smaller range in the vertical scale). There is a strong correlation; a slope of $0.21 \pm 0.04$ and rms dispersion of 0.05 mag. This figure confirms that the $NUV$ scatter is not due to contamination by optically blue galaxies.

The analogous $NUV$--$J$ correlation with the metallicity is given in Figure \ref{fig:met}. There is a strong trend between [Z/H] and $NUV$--$J$, with a slope of $1.27 \pm 0.23$ and an rms dispersion of 0.32 mag ($\sim$90 per cent of which is intrinsic scatter). The lack of galaxies to the top left is not the result of a selection effect. Of the two target galaxies undetected in the $NUV$ band (see Sec. \ref{sec:compare}), only one has a low metallicity. Assuming the non-detection is due to a redder-than-average colour, this would add a single galaxy to the upper left of the plot, but would not significantly affect the fit. In general, the most luminous galaxies form a ridgeline at redder colours, although there are significant bright outliers to this trend (the most obvious being the bright, blue, metal-rich galaxy on the right; NFPJ132729.7--312325). Lower metallicity galaxies tend to be bluer and less luminous. This supports \citet{ram07}, who use simulations to predict a correlation of $NUV$--IR colour with metallicity, but little dependence on age in populations greater than 2--3 Gyr after a star formation episode. However, the slope of our observed metallicity trend is 2--3 times weaker than that derived from theoretical spectra by \citet{dor03}.

Model evolutionary tracks suggest that stellar evolution depends on $\alpha$/Fe \citep{sal00,dot07}. However, we find no discernible relation between the $NUV$--$J$ colour and the $\alpha$-abundance in this sample of red sequence galaxies (rms scatter of 0.37 mag; Fig. \ref{fig:alphafe}).

\begin{figure}
\includegraphics[viewport=0mm 0mm 100mm 134mm,height=84mm,angle=270,clip]{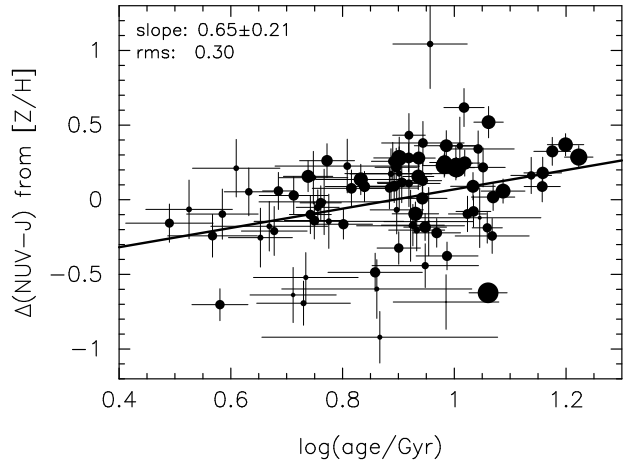}
\caption{$NUV$--$J$ residuals from the [Z/H] relation vs log(age). Symbol size represents $J$ band luminosity; larger=brighter.}
\label{fig:age_resid}
\end{figure}

Figure \ref{fig:age_resid} shows the residuals from the $NUV$ vs metallicity relation (Fig. \ref{fig:met}) against log(age). Although the residuals are more strongly correlated with age ($\sim$$3\,\sigma$) than the colours themselves are, the rms dispersion is only reduced to 0.3 mag ($\sim$0.25 mag intrinsic scatter).  This is in contrast to the case of the $B$-$R$ colour where the majority of the scatter can be attributed to age and metallicity \citep{smi08}.

\section{Discussion}
\label{sec:discuss}

This section explores possible causes of the large intrinsic scatter observed in the $NUV$--$J$ colour. We investigate aperture bias, the UV upturn phenomenon and `frosting' by young or metal poor subpopulations. Additionally, we comment on the uncertainty in the $NUV$ band K-corrections.

\subsection{Aperture bias/morphology}
The spectroscopy used to estimate the stellar parameters is derived from 2 arcsec diameter fibres, whilst the $NUV$--$J$ colours are from 12 arcsec diameter aperture photometry. Eliminating aperture bias completely would require matched apertures for the photometry and spectroscopy, but unfortunately the PSF of the 2MASS and {\em GALEX} images is too large for reliable 2 arcsec aperture photometry. 

Broadband colours and spectroscopic measurements (e.g. \citealt{tam04}, \citealt{san07}) show that elliptical galaxies generally have flat radial profiles in age, and regular metallicity gradients (decreasing [Z/H] with radius). The aperture effect therefore flattens the $NUV$--$J$ vs Z/H relation, as larger, redder galaxies will tend to exhibit higher metallicities within the fibre. However, the effect is small, with only $\sim$0.1 dex change in metallicity over a 1 dex difference in aperture radius.

\begin{figure}
\includegraphics[viewport=49mm 60mm 143mm 182mm,height=84mm,angle=270,clip]{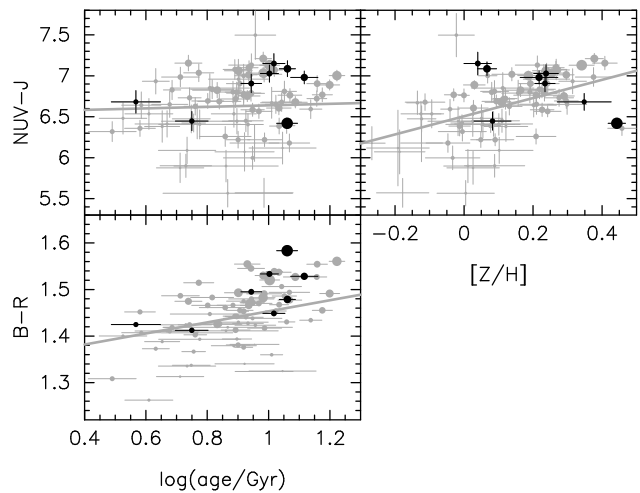}
\caption{Colour vs stellar parameters, highlighting galaxies with available {\em HST}-ACS images. Symbol size reflects $J$ band luminosity (larger=brighter). In all cases, ACS confirms early-type morphology, uncontaminated by neighbours.}
\label{fig:withmorph}
\end{figure}

\begin{figure}
\centering
\includegraphics[width=55mm]{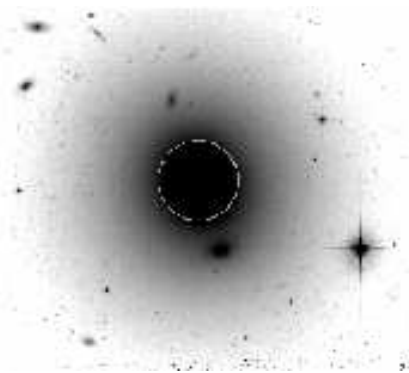}
\caption{{\em HST}-ACS image of the outlier NFPJ132729.7--312325, with a 12 arcsec diameter aperture marked.}
\label{fig:obj}
\end{figure}

Inspection of the galaxies in high resolution images can ascertain whether there are morphological peculiarities, or neighbouring objects, contributing to an enhanced large radius $NUV$ flux. Figure \ref{fig:withmorph} shows colour vs age and metallicity (as in Figs. \ref{fig:age}--\ref{fig:met}), highlighting the few galaxies that have high resolution {\em HST}-ACS images available. Fortunately, one of these objects is the most obvious outlier in the whole sample (NFPJ132729.7--312325); it is the reddest in $B$--$R$, but has an unusually blue $NUV$--$J$ colour for such a metal-rich, luminous galaxy. In the {\em HST}-ACS image (Fig. \ref{fig:obj}), the galaxy appears to be a large elliptical with no abnormalities. There are also no obvious contaminating objects that could be responsible for an anomalous blue colour. 

On close examination, the other targets with {\em HST}-ACS images also appear to be `normal' ellipticals of various sizes, with no obvious peculiarities or large radius $NUV$ contributors. We conclude that a large scatter in $NUV$--$J$ colour is present even in objects with confirmed regular early-type morphologies and no contaminants. Additionally, a scatter of $\sim$0.3--0.35 mag is still obtained when smaller matched apertures (4.5 arcsec diameter) are used for the colour, despite the probability of a contaminating neighbour being reduced by $\sim$85 per cent.

\subsection{UV upturn}
\label{sec:uvx}
Another possible explanation for the scatter in $NUV$--$J$ colour is $NUV$ contamination by the UV upturn (UVX). The \citet{tho03} models used in the stellar populations parameter calculations include low-metallicity blue horizontal branch sub-populations and thermal pulsing asymptotic giant branch stars, but they do not include the low mass, metal-rich, helium burning stars with small envelopes currently thought to be the most likely candidate for the UV upturn \citep{oco99}. The UV upturn is one of the most heterogeneous photometric properties of old stellar populations in early-type galaxies, with a spread of up to $\sim$4 magnitudes in the $FUV$ \citep{oco99}, so certainly seems a plausible explanation for the scatter.

The hot UVX component appears in the spectra as a smooth continuum with an absence of emission and absorption lines. Hence, the $FUV$ flux from a galaxy can be used to estimate the extent of the contribution in the $NUV$ band \citep{bur88}. Unfortunately, none of the SSC galaxies have available $FUV$ photometry, so in order to estimate the extent of the UVX contribution, the corrections are calculated for all objects in the NFPS sample that have the necessary $FUV$ data. However, stellar population ages have not been derived for individual NFPS galaxies, so instead we compare the corrected $NUV$--$J$ colours to the traditional `age-tracing' Balmer line H$\gamma$F. This line does not trace age cleanly, being affected to a small degree by the metallicity.

\begin{figure}
\includegraphics[viewport=0mm 0mm 177mm 125mm,height=84mm,angle=270,clip]{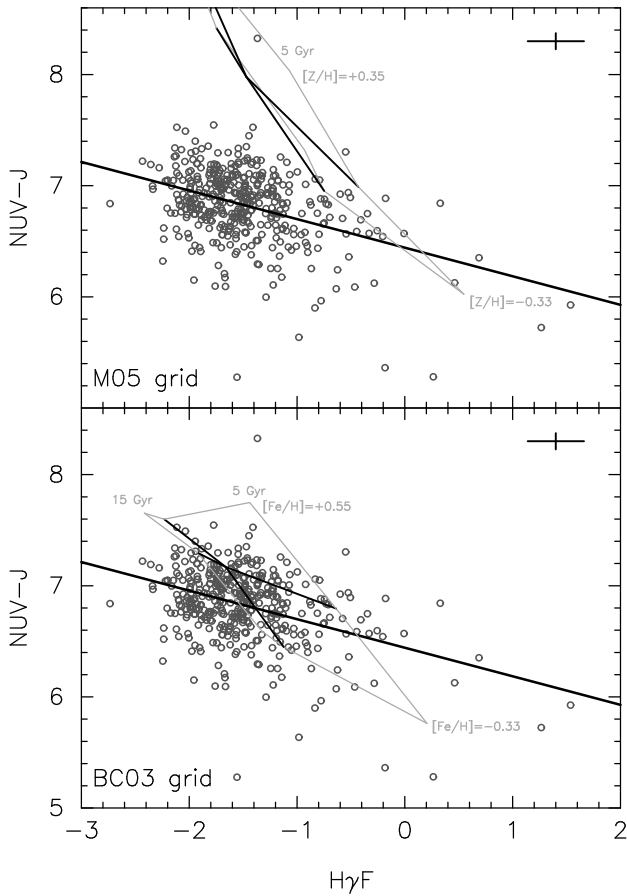}
\caption{$NUV$--$J$ (without K-correction) against H$\gamma$F Balmer index for NFPS galaxies. Upper panel: M05 grid (age=\{5,12,15 Gyr\}, [Z/H]=\{--0.33,0.00,+0.35\}). Lower panel: BC03 grid (age=\{5,12,15 Gyr\}, [Fe/H]=\{--0.33,+0.09,+0.55\}). (5,--0.33) is the bluest (age,[Z/H]) grid point. Median error bars are shown top right.}
\label{fig:nfps_balmer}
\end{figure}

H$\gamma$F vs $NUV$--$J$ for the NFPS sample is shown in Figure \ref{fig:nfps_balmer}. Given the relative strengths of the colour vs. age and [Z/H] relations (Figures \ref{fig:age} and \ref{fig:met}), the correlation seen here is most likely a reflection of the metallicity, rather than age, dependence of H$\gamma$F. For comparison, the observations have been overlaid by age/metallicity grids constructed from the models of \citet[M05; upper panel]{mara05} and \citet[BC03; lower panel]{bru03}. Both grids lie redward of the observed data, most likely due to neither model including EHB stars, although the BC03 grid provides the better description of the data. Previous studies (e.g. \citealt{sal07}) have noted that while the BC03 models do not explicitly include EHB stars, the UV light from old stellar populations (primarily post-AGB stars) reproduce several of the correlations found in observational data, including the relation between $FUV$--$NUV$ and $B$--$V$ colours \citep[see][particularly fig. 5]{don06}.

\citet{dor03} introduced corrections to their $NUV$--$V$ colours by considering the relative contributions of the hot and cool (i.e. non-UVX) components at $FUV$, $NUV$ and $V$. They assumed a negligible contribution of the cool component to the $FUV$ flux, but allowed hot stars to contribute at $V$. Here, we use analogous corrections for the $NUV$--$J$ colours, which are simpler because it is safe to assume no UVX contribution at $J$. The corrected colours are given by  
\begin{eqnarray}
(NUV-J)_{\rm corr} = -2.5\times log[10^{-0.4(NUV-J)_{\rm obs}} \nonumber \\
       -\alpha\times10^{-0.4(FUV-J)_{\rm obs}}]
\end{eqnarray}
where $\alpha\approx0.3$ is the ratio of $NUV$ to $FUV$ flux for the hot component (appropriate for a $T_{\rm eff}=24000$\,K star, see Dorman et al.).

\begin{figure}
\includegraphics[viewport=49mm 26mm 147mm 150mm,height=84mm,angle=270,clip]{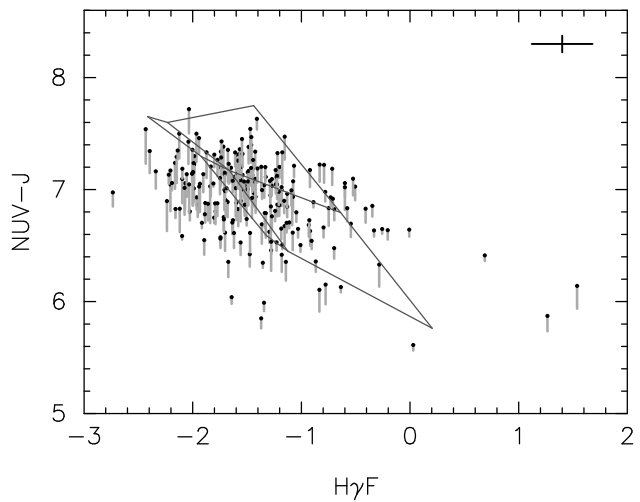}
\caption{$NUV$--$J$ (without K-correction) against H$\gamma$F Balmer index for galaxies with available $FUV$ data. Points indicate the corrected colour, with the `tail' showing the correction vector. BC03 grid as in Fig. \ref{fig:nfps_balmer} (lower panel). Median error bars are shown top right.}
\label{fig:uvx_corr}
\end{figure}

The resulting corrections are shown as vertical lines in Figure \ref{fig:uvx_corr} (where the points indicate the value of the corrected colour). The corrections move the colours redward as expected, but are only of the order of $\sim$0.2 magnitudes, and do not reduce the scatter in the $NUV$--$J$ (0.29 mag before corrections, 0.30 mag after). Obviously there are no extreme UVX galaxies with large hot component contributions to the $NUV$ flux in this sample, and as such the UVX phenomenon is unlikely to responsible for the scatter in $NUV$--$J$ for NFPS. We speculate that the UVX effects in the SSC sample are smaller because, on average, the galaxies have lower luminosities.

\subsection{Stellar population `frosting'}
\label{sec:frost}
The simple models can be generalised by constructing composite stellar populations. A common invocation of this is residual star formation in the form of `frosting' galaxies with a small mass fraction of young stars \citep{tra00}, which manifests itself in the observables as a bluer colour and a younger age than the base population. Frosting should not affect the spectroscopic metallicity or $\alpha$-abundance significantly as these are primarily driven by the larger, older population \citep{ser07}.

\begin{figure}
\includegraphics[viewport=0mm 0mm 100mm 130mm,height=84mm,angle=270,clip]{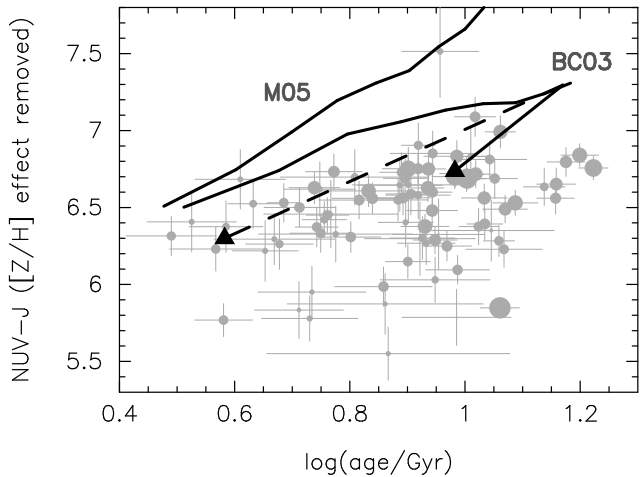}
\caption{Simple stellar population tracks from the M05 and BC03 models shown in the context of the $NUV$--$J$ colour (correcting to [Z/H]=0 using the correlation from Fig. \ref{fig:met}) vs log(age). Vectors indicate two BC03 frosting scenarios with minor populations of  \{$\mu$,age\} $=$ \{0.03,1.5Gyr\} and \{0.01,0.7Gyr\} (solid and dashed vector respectively).}
\label{fig:frosting}
\end{figure}

Figure \ref{fig:frosting} shows the extent to which frosting can account for the scatter, after the effect of metallicity has been removed using the correlation of Fig. \ref{fig:met}. Simple stellar population (SSP) tracks from both the M05 and BC03 models are shown. The slope of the BC03 track fits the observed red envelope well, whereas M05 predicts a much steeper variation with age. Allowing for the systematic effects of aperture bias and the UV upturn ($\sim$0.3 mag bluer in observed $NUV$--$J$), $\sim$30 per cent of the galaxies cannot be accounted for by the SSP model.

Vectors of the frosting effect on $NUV$--$J$ colour and spectroscopic age have been calculated using the BC03 models. Frosting of a 15 Gyr base population by a 1.5 Gyr population with a mass fraction $\mu$=0.03, and by a 0.7 Gyr population of $\mu$=0.01 are shown. The M05 models result in marginally steeper frosting vectors, as would be expected given the steeper SSP track. The vectors show that frosting, even at a modest level of 1--2 per cent for a $\sim$1 Gyr population, could account for a sizeable portion of the scatter in the $NUV$--$J$ colour. A young population of this size would not be apparent in the $B$--$R$ colours.

The spectroscopic age is sensitive to frosting via the increased hot-star contribution to the Balmer lines, and for a given change to the spectroscopic age, the UV colours are affected more strongly by
frosting than by lowering the age of a single-burst population. This supports the assertion that UV colours are partly dependent on low level recent star formation (\citealt{fer00}, \citealt{kav06}, \citealt{sal07}). However, there are spectroscopically old galaxies with blue colours ($\sim$10 per cent of the sample) which cannot be accounted for by the frosting scenario described above. Additionally, Rose CaII index results \citep{smi08} appear not to support the presence of young stellar populations ($\la$ 1.5 Gyr) in the majority of red sequence galaxies.

For most of this study, we have neglected the blue horizontal branch (BHB) as cluster red sequence galaxies have [Z/H] $\sim$0. However, it is possible that a low metallicity population with a BHB morphology may be present in some galaxies \citep{mara00}. Therefore, an alternative `frosting' scenario consists of a low mass fraction, old, low metallicity, BHB stellar population embedded in a [Z/H]=0 galaxy. Estimates from the M05 models show that `frosting' by a 4--5 per cent mass fraction population with [Z/H]=--1.35 could give an $NUV$--$J$ colour $\sim$1 mag bluer, while changing the $B$--$R$ by only $\sim$0.03 mag. The effect of the low-metallicity frosting on the derived spectroscopic age is likely similar to the effect of frosting by a young component, since both cases are driven by increased A-star contribution to the Balmer lines. 

\subsection{$NUV$ K-corrections}
\label{sec:kcor}
K-corrections for the UV bands are poorly constrained, as the spectral shape of galaxies at wavelengths shorter than $\sim$3000\AA { }is not well known. Previous studies of the UV CMR largely ignore the K-correction, or in the case of \citet[who estimate corrections of $0.1-0.2$ mag for $z = 0-0.25$]{yi05} apply correcions based on the luminosity distance without considering the spectral shape of galaxies. 

\citet{kav06} compute $NUV$ K-corrections for best fit model SEDs derived from SDSS and {\em GALEX} photometry. Corrections of $\sim$0.1 mag were found for redshifts z $<$ 0.1. However, corrections for their model of a 9 Gyr old simple stellar population are much larger ($0.4-1.0$ mag for 0.04 $<$ z $<$ 0.11).

As an illustration, we calculate K-corrections from the models of \citet[M05]{mara05} at various redshifts (0.015 $<$ z $<$ 0.072), both for simple populations and galaxies in the `frosting' scenario described in Sec \ref{sec:frost}. K-corrections of 0.2--1.2 mag are obtained, depending on the metallicity and age (increasing either increases the correction), and on the mass fraction of the young stellar component. The corrections are dominated by the 2640\AA{ }spectral break (cf \citealt{eis03}), which is redshifted completely out of the $NUV$ band by $z \approx 0.07$. In \citeauthor{eis03}, who study the average spectra of 726 luminous, red, SDSS galaxies at $0.47 < z < 0.55$, the break appears less prominent than in the M05 spectra (a $\sim$50 per cent drop in flux as opposed to $\sim$80 per cent).

\begin{figure}
\includegraphics[viewport=40mm 50mm 140mm 185mm,height=84mm,angle=270,clip]{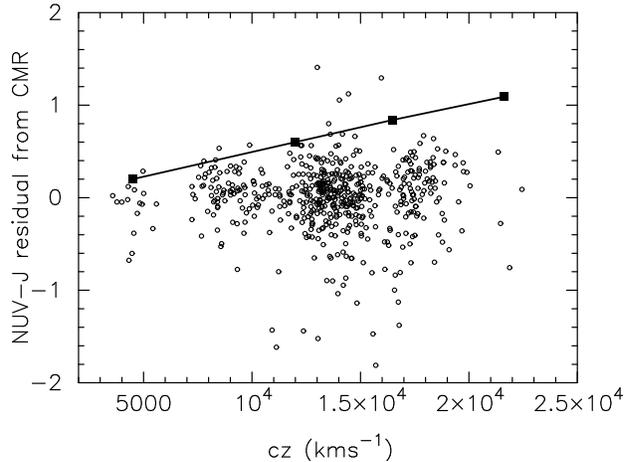}
\caption{CMR residuals for the NFPS sample (Fig \ref{fig:cmdiag}; upper panel) vs heliocentric galaxy redshift. The filled squares and solid line show the predicted trend if the K-corrections from the M05 models were necessary, but not applied.}
\label{fig:kcor}
\end{figure}

Fig \ref{fig:kcor} shows the regression line residuals from the colour--magnitude diagram for the NFPS sample (Fig \ref{fig:cmdiag}; upper panel) plotted against the galaxy redshift in the heliocentric frame. The solid line shows the expected trend if the M05 K-corrections were necessary, but not applied. This implies the uncorrected residuals should have a steep correlation with redshift, which is not observed.  A flatter K-correction is preferred by the data, as in \citet{kav06}. Along with the \citeauthor{eis03} spectrum described above, this result highlights the uncertainty in the UV K-corrections (as well as the problems of stellar population models in the UV).

\section{Conclusions}
\label{sec:conclusion}
Using {\em GALEX} UV and 2MASS $J$ band photometry, we have investigated the relationship between UV--IR colours and spectroscopically derived stellar population parameters (age, metallicity and $\alpha$-abundance) for red sequence galaxies in local clusters.

We select galaxies using strict emission criteria to avoid contamination from galaxies with very recent star formation. We analyse the $NUV$ colour--magnitude relation (CMR) for our two samples of quiescent galaxies (920 in NFPS; 156 in SSC), and find rms dispersions of 0.37 and 0.30 mag (intrinsic scatter of 0.36 and 0.29 mag) respectively. This is similar to previously reported values of $\sim$0.5 mag and is an order of magnitude larger than the scatter in the optical CMR ($\sim$0.05 mag).

We compared the $NUV$--$J$ colour to the spectroscopic stellar population parameters for 87 galaxies in the SSC sample and found the following:

\begin{itemize}
\item There is a significant $NUV$--$J$ vs metallicity trend, with a slope of $1.27 \pm 0.23$ and an rms dispersion of 0.32 mag.
\item There is only a weak $NUV$--$J$ vs age trend after the metallicity effect has been removed, and no correlation with $\alpha$-abundance.
\item There is a large intrinsic scatter ($\sim$0.25 mag) in the $NUV$--$J$ colour at fixed age and metallicity which cannot be easily accounted for with simple stellar populations.
\end{itemize}

The unexpected blue colours of at least some objects, including an influential outlier, cannot be attributed to large radius contamination from other objects, and aperture bias cannot account for the large scatter. Corrections for the UV upturn (UVX) phenomenon are relatively small ($\sim$0.2 mag) and are similar galaxy-to-galaxy, so do not reduce the intrinsic scatter.

We find that the large $NUV$--$J$ intrinsic scatter could be attributed to galaxy `frosting' by small ($<$ 5 per cent) populations of either young stars or a low metallicity blue horizontal branch.

\section*{Acknowledgments}
TDR is supported by the STFC Studentship PPA/S/S/2006/04341. RJS is supported by the rolling grant PP/C501568/1 `Extragalactic Astronomy and Cosmology at Durham 2005--2010'. 
Based on observations made with the NASA Galaxy Evolution Explorer. {\em GALEX} is operated for NASA by California Institute of Technology under NASA contract NAS-98034. This publication makes use of data products from the Two Micron All Sky Survey, which is a joint project of the University of Massachusetts and the Infrared Processing and Analysis Center/California Institute of Technology, funded by NASA and the National Science Foundation. We thank Chris Haines and the SOS team for providing the $B$--$R$ colours used in Figures \ref{fig:age} and \ref{fig:withmorph}.

\label{lastpage}

\end{document}